\documentclass{phbauth}
\usepackage{graphicx}

\begin{document}

\begin{frontmatter}

\title{Superconductivity, magnetic ordering, and its interplay in 
HoNi$_2$$^{11}$B$_2$C}

\author[address1]{A. Dertinger\thanksref{thank1}},
\author[address2]{A. Kreyssig},
\author[address3]{C. Ritter},
\author[address2]{M. Loewenhaupt}, and
\author[address1]{H.F. Braun}

\address[address1]{Physikalisches Institut, Experimentalphysik V, 
Universit\"at Bayreuth, D-95440 Bayreuth, Germany}

\address[address2]{Institut f\"ur Angewandte Physik, Technische 
Universit\"at Dresden, D-01062 Dresden, Germany}

\address[address3]{Institut Max von Laue - Paul Langevin, 156 X, F-38042 
Grenoble, France}

\thanks[thank1]{Corresponding author: armin.dertinger@uni-bayreuth.de}

\begin{abstract}
We investigated the low-temperature properties of polycrystalline 
HoNi$_{2}$$^{11}$B$_{2}$C by means of ac-susceptibility and neutron powder 
diffraction in the temperature range of 1.5 to 10~K and zero magnetic field. 
A sample with pronounced re-entrant behaviour was chosen since it exhibits a 
superconducting state which is strongly affected by the intrinsic magnetism, 
the formation of long range magnetic order, via pair-breaking. The evolution 
and volume fractions of the three different magnetic structures were observed 
and correlated to the superconducting behaviour.
\end{abstract}

\end{frontmatter}

\section{Introduction}

In the family of the quaternary borocarbides {\it R\/}Ni$_2$B$_2$C, 
co-existence of superconductivity and antiferromagnetism of the rare earth 
ions has been found at low temperatures for {\it R\/} = Tm, Er, Ho, and Dy 
(see e.g. \cite{Lynn97,Canfield98}). In the case of HoNi$_2$B$_2$C, the 
transition temperatures for both ordering phenomena lie in the same 
temperature region of about 5 to 8~K. Hence, this system offers an ideal 
possibility to study the interplay between magnetism and superconductivity. 
Three different magnetic structures have been identified for HoNi$_2$B$_2$C: a 
commensurate antiferromagnetic one with wave vector $\tau_1 = (0,0,1)$, an 
incommensurate structure with $\tau_2 = (0,0,0.915)$ and an incommensurate 
structure with $\tau_3 = (0.585,0,0)$ \cite{Lynn97,Kreyssig97}. While for 
flux-grown single crystals only an ordinary superconducting transition has 
been observed in zero magnetic field, the low-temperature properties of 
polycrystalline samples strongly depend on thermal treatment \cite{Schmidt97} 
as well as on their chemical composition within a small but non-negligible 
homogeneity range \cite{Schmidt98}. Pure superconductivity co-existing with 
antiferromagnetism, re-entrant behaviour, as well as samples in a 
non-superconducting but magnetically ordered state down to at least 1.5~K have 
been found \cite{Schmidt96}. The study of different polycrystalline samples 
can therefore reveal correlations between the complicated set of magnetic 
structures and the occurrence of superconductivity. In this paper, we 
will present results on a sample with pronounced re-entrant behaviour in zero 
magnetic field.

\begin{figure}[t]
\begin{center}\leavevmode
\includegraphics[width=0.8\linewidth]{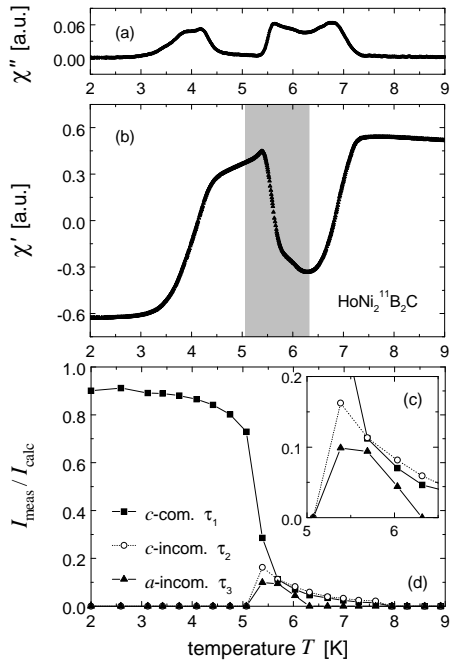}
\caption{Temperature dependence of the real and imaginary part of the 
ac-susceptibility $\chi'$ (b) and $\chi''$ (a) and of the intensities of the 
magnetic reflections $\tau_1$, $\tau_2$ and $\tau_3$ (d), corresponding to the 
commensurate $c$-axis, the incommensurate $c$-axis and the incommensurate 
$a$-axis modulated structure, respectively. The inset (c) shows an enlargement 
of the intensities of $\tau_2$ and $\tau_3$. All intensities are normalized by 
the calculated intensity for the fully magnetically ordered phase with the 
free ion value of $10\ \mu_{\rm B}$ for Ho$^{3+}$. The shaded area in (b) 
corresponds to the temperature interval where the $a$-incommensurate 
structure exists.
}\label{LTFarmin}\end{center}\end{figure}

\section{Experimental}

The HoNi$_{2}$$^{11}$B$_{2}$C sample was prepared by arc-melting 
stoichiometric amounts of the pure elements on a water-cooled copper hearth 
under argon atmosphere. 10~at.-\%\ excess carbon were added to compensate for 
evaporation losses. The sample was remelted several times and then annealed 
for 16~days at 1100$^\circ$C to ensure its homogeneity. Phase purity was 
checked by X-ray powder diffraction. Only minor traces of HoB$_2$C$_2$ could 
be detected. The ac-susceptibility measurements were performed at 22~Hz with 
an ac-amplitude of 0.1~mT. The neutron powder diffraction experiment was 
carried out at the multidetector instrument D1B at the ILL in Grenoble.

\section{Results and discussion}

The ac-susceptibility measurements can be seen in Fig.\,\ref{LTFarmin}\,b. The 
sample becomes superconducting at 7.1~K, re-enters the normal state below 
roughly 6.3~K, and finally becomes superconducting again at about 4.3~K. Each 
transition between superconducting and normal state, and vice versa, is 
accompanied by a loss signal in the quadrature part of the susceptibility 
$\chi''$ (Fig.\,\ref{LTFarmin}\,a). 

The neutron scattering data in Fig.\,\ref{LTFarmin}\,d show first signs of 
long range magnetic ordering with the appearance of the two $c$-axis modulated 
structures at about 8~K. Their tail exists well above the superconducting 
transition temperature whereas the $a$-axis modulated structure is confined to 
a very small temperature interval below $T_c$ (see Fig.\,\ref{LTFarmin}\,c). 
This temperature region is highlighted by shading in the susceptibility plot 
and is exactly where the sample re-enters the normal state. It indicates that 
this structure is responsible for the observed pair-breaking which leads to 
re-entrant behaviour. This interpretation is in very good agreement with 
neutron scattering experiments on samples with non-magnetic dilutions on the 
Ho-site \cite{Kreyssig97,Kreyssig98}. However, the normal state persists below 
the temperature where the $a$-incommensurate structure vanishes. Below 5~K, 
only the $c$-commensurate groundstate is observed and superconductivity 
returns at 4.3~K.

We gratefully acknowledge the assistance of W. Ettig and A. Krause. This work 
was partially supported by the DFG through SFB 463.

\end{document}